# Influence of probe geometry of optical fibers in sensing volatile liquids through localized surface plasmon resonance

D. Paul[a], R. Biswas[a]* and B. S. Boruah[a]

*Abstract*— A comprehensive analysis of influence of probe geometry on localized surface plasmon resonance phenomenon is presented. Plasmonic responses of noble metal nanoparticles such as AuNPs and AgNPs have been exclusively examined by adopting D-type and tapered optical fiber probes for the detection of methanol. With increase in concentration of methanol, the effective refractive index of the medium changes which leads to alter absorbance characteristics of the NPs; thereby modulating the eventual output responses of the proposed sensor. A comparative study has been presented with respective change in geometrical shapes of the probes to detect methanol. The sensitivity in case of D-type probe for detection of methanol is found to be ~0.09644 mV/ppm with AuNPs and ~0.03038 mV/ppm with AgNPs. On the other hand, the sensitivity for AgNPs coated probe is found to be ~0.00389 mV/ppm and 0.00379 mV/ppm for AuNPs in case of tapered probe.

## I. INTRODUCTION

Optical fiber probe provides miniaturized platform as compared to bulky prism and microscopic system for plasmonic responses [1-5]. The use of an optical fiber probe makes a sensing system simple, compact and cost-effective as well as portable [3]. The performance of a sensor can be improved by utilizing optical fiber probes of variable geometries [2-3]. Till date, a number of optical fiber sensors have been reported with variable geometries such as tapered, D-type, U-bent, hetero-core structured, arrayed fiber end face and so on [2-9]. Change in probe geometry affects the coupling of light with analytes and thereby alters distribution of evanescent wave onto the exposed portion of a probe [3]. Fiber optic sensor based on localized surface plasmon resonance (LSPR) phenomenon is one of the powerful tools for label free sensing and point of care applications [10-12]. Utilization of noble metal NPs aids control and optimization in sensing system [3-5]. Moreover, plasmonic response of noble metal nanoparticles (viz. gold and silver NPs) can also be tuned by varying the geometry of a probe in a fiber-optic sensing system. Of late, extensive research has been carried out by adopting LSPR enhanced U-bent optical fiber probe based sensing of different volatile liquids, size of NPs as well as detection of heavy metal ions [6-15]. Different research groups have demonstrated techniques to enhance the sensitivity of the evanescent wave across the unclad region of the fiber by modifying the probe into different shapes such as tapered, tapered tip, biconal taper, straight and U-bent [7-17]. However, to the best of our knowledge, low cost LSPR based volatile liquid sensing using diverse probe geometries has not been addressed till now. As a proof of concept, two variable geometrical fibers have been utilized to detect the proposed analyte along with a comparative study between each shape.

This present work is aimed at outlining of the sensing capabilities of D-type and tapered optical fiber probes with respective change in effective refractive index of the used analyte i.e., methanol. In a D-type optical fiber probe, portion alike D shape of the cladding (~1cm) is removed at the middle whereas in tapered optical fiber probe, the whole central cladding (~1cm) part is removed through chemical etching technique. The tapered one has been fabricated by heating and pulling technique. Schematic of D-shaped as well as tapered fiber probe is illustrated in Fig. 1(a). In the tapered fiber, the mid portion is meant for impregnation with NPs. Likewise, in D-shaped fiber too, the mid D-shaped region is meant for impregnation. Each probe has been fabricated according to feasibility that can offer minimal loss. Utilizing aforementioned probes, methanol has been used as analyte with coating of noble metal NPs onto the sensing region. Further, a low cost optical detector (CMOS) has been used to observe the responses in lieu of an expensive spectrophotometer. The use of a simple detector not only reduces the cost of the set-up but also makes the set-up compact.

## II. DETAILS OF EXPERIMENTAL SET-UP

The proposed sensing set-up is composed of a test chamber where the probes are fixed individually while checking the responses. After fabricating the probe with heating as well as pulling technique [3], the exposed portion is cleaned and impregnated with each noble metal NPs, independently. A broadband light source [Thorlabs, SLS202L] has been used which is followed by a collimator [Thorlabs, F810FC], and on the other side of the set-up, an optical detector [Coherent Optics, 1098313 RoHS] is affixed to observe the changes via a collimator. Methanol is kept inside test chamber. The schematic is depicted in Figure 1(b). While performing the experiment, a multimode silica optical fiber has been used having core/cladding diameter of 800/1000 μm and numerical aperture (NA) ~0.39 [Thorlabs], which has been used for fabricating the probe for experiment. For detection purpose, a CMOS (metal-oxide: $SnO_2$) sensing head has been used during the experiment along with microcontroller (Arduino Inc.). The sensing head is a low cost semiconductor sensor; capable of detecting the presence of volatile liquids of our designed sensor. Thus, the variation in concentration of volatile liquid has been observed using the designed sensing head.

[a]Applied optics and photonics laboratory, Department of Physics, Tezpur University, Napaam Sonitpur-784028 (First author e-mail: dmppaul22@gmail.com)
*Corresponding author e-mail: rajib@tezu.ernet.in

The noble metal NPs particles of size ~40nm are synthesized following [20-24], by reduction of trisodium citrate ($Na_3C_6H_5O_7 \cdot 2H_2O$) method. The shape and size of each NP's have been characterized by UV-Vis and TEM analysis. Also, to coat the NPs onto the decladded portions, each fiber has been kept into piranha solution for about 10 to 15 minutes, followed by acetic acid for about 2 to 3 hours for exterior silanization. The aforementioned treatment onto the decladded portions of the fibers leads to self-assembly of NPs which helps to detect the proposed analyte [19].

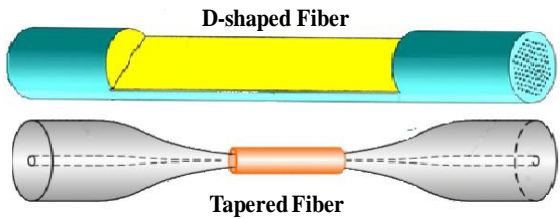

Fig. 1 (a) Schematic of D-shaped and tapered fiber probe

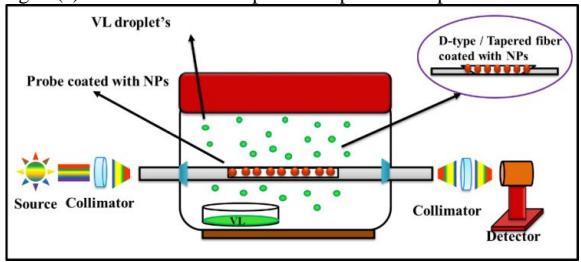

Fig. 1 (b) Schematic of the set-up

### III. RESULTS AND DISCUSSIONS

The probe geometry can alter the sensing performance of a sensing set-up. The evanescent wave being coupled with noble metal NPs resides on the exposed portion of the probe that brings measurable changes as detected by the optical receiver. Injection of VLs near to the NPs coated probe leads to vary the effective RI of the adjacent environment according to LSPR phenomenon. Accordingly the variation has been observed in the output with respect to the proposed probes. For D-type probe coupling occur only the half of the cylindrical exposed portion. In tapered probe, coupling hovers around the beam waist. Thus, with progressive change in concentration of the proposed VL, the responses can be changed which thereby facilitates detection of the proposed VL.

To calculate concentration, a sensing head along with microcontroller (Arduino Inc.) device has been used. The sensor can also easily detect the change in vapor concentration in real-time as illustrated by Fig. 1 (c). In order to attain the response from the sensing head, the concentration of volatile liquid has been changed in known concentrations and the corresponding output has been taken. The calibrated curve (concentration as abscissa and voltage response as ordinate) in Fig. 1 (c) shows the changes of the VLs inside the chamber. The initial output of the sensing head has been recorded to be ~0.32 mV without introduction of VL and is found to be increasing with rising concentration of volatile liquid, accompanied by a eventual saturation stage of ~6.7 mV nearing 600 ppm. Also, SEM image of coating of AuNPs on D-shaped fiber as well as TEM images of the NPs have been provided as Fig. 1 (d)-(f).

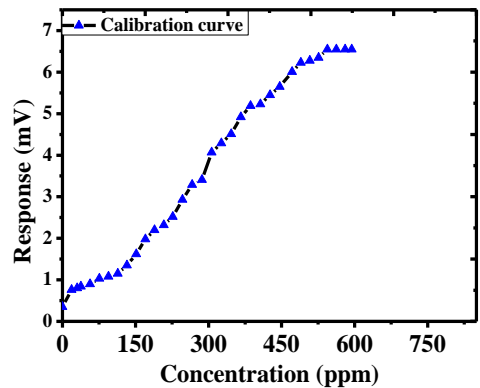

Fig 1 (c). Calibration curve of the sensing head

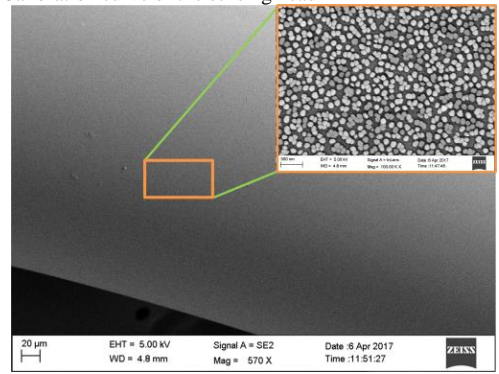

Fig. 1 (d) SEM image of the exposed portion of the D-shaped fiber coated with AuNPs.

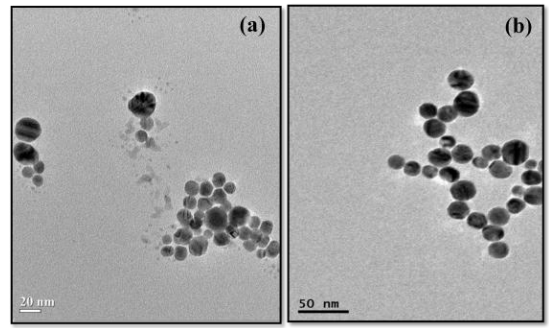

Fig. 1 (e) TEM images of the 40nm NPs (a) AuNPs and (b) AgNPs

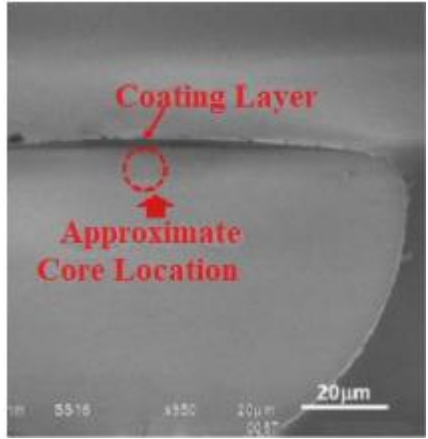

Fig. 1 (f) SEM image of AuNPs coated layer in case of D-shaped fiber

The purpose of the paper is to compare sensitivity of noble metal NPs in presence of methanol with variable geometries of the probes (i.e., D-type and tapered). The responses in each set of fiber probe have been measured in terms of voltages with corresponding change in concentration of the proposed volatile liquid. Optical response of each probe has been observed by impregnation of the NPs onto the exposed portion with suitable technique. Blank response for each set of probes of the sensing set-up with each NP's has been noted before initiating the experiment. It is to be mentioned that D-type optical fiber allows the coupling to only half of the cylindrical portion [Fig1 (a)]. Initially, in absence of methanol, the response of the sensing set-up is found to be 2.2 mV for AuNPs and 2.1 mV for AgNPs probe, respectively. After introduction of methanol onto the test chamber, the response of the set-up starts increasing with increase in concentration of the proposed analyte. Volatile nature of methanol leads to change in the effective refractive index of the medium adjacent to the NPs of the environment inside the test chamber. With passage of time, the concentration gradually rises which accordingly increases the effective refractive index of the medium. As such, it modulates responses of sensing set-up. The responses of each noble metal NPs for D-type probe are illustrated in figure 2. There is a dip observed in each set of NPs coated in case of D-type probe which is ascribed to the coupling of evanescent wave field with the half of the exposed portion of the fiber. In D-type probe, only half of the probe is exposed; while the other half is utilized by the electromagnetic wave to complete the total internal reflection (TIR). This leads to attenuated TIR. This attenuated TIR prevails until the beam reaches at the core-cladding interface which leads to dip in the output. Thus, the variation has been observed in the output for AuNPs and AgNPs coated probe in presence of methanol. Eventually, there occurs saturation at 27.02 and 16.89 mV for AuNPs and AgNPs, respectively once complete vaporization of methanol inside the chamber takes place. In the working domain of 0-235.37 ppm, the sensitivity of the sensing set-up is found to be 0.0964 mV/ppm with limit of detection (LOD) of $174.24\times10^{-2}$ ppm for AuNPs coated probe. Table 1 enlisted the performances of the sensor for D-type probe in case of AgNPs coated probe respectively.

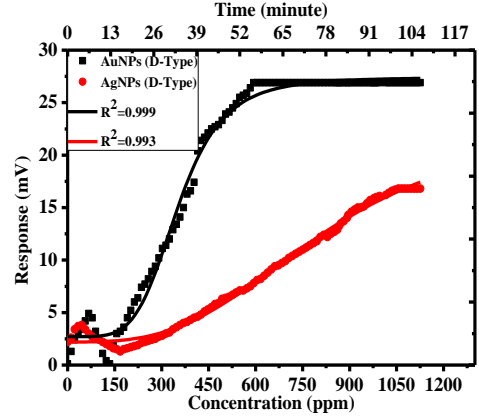

Fig 2. Response of the sensing set-up vs. concentration of methanol in case of AuNPs and AgNPs coated probe

As observed in figure 2, the set-up follows sigmoid pattern and thus, from fitting, the regression in each case of noble metal NPs is found to be ~1. The empirical equations of D-type probe followed by the AuNPs and AgNPs coated probes are as follows:

AuNPs: $Y_m = 27.398 - \dfrac{24.69}{1+\left(\dfrac{X_m}{352.09}\right)^{4.9}}$  1 (a)

AgNPs: $Y_m = 23.2496 - \dfrac{21.053}{1+\left(\dfrac{X_m}{824.15}\right)^{3.01}}$  1 (b)

Here, $Y_m$ (dependent parameter) and $X_m$ (independent parameter) represent voltage and concentration, respectively. Likewise, the variation has been observed for tapered optical fiber probe. Before initiating the experiment, the probe has been cleaned and dried by keeping the chamber free from volatile liquid. To confirm that the chamber is free from methanol, the initial reading has been noted for a while and after getting a fixed value, experiment has been initiated by gradual increase in concentration of analyte. Initially, with the tapered probe, the response is found to be 0.5 mV and 0.44 mV for AuNPs and AgNPs, respectively. Since, tapered fiber probe is fully decladded at the center; the evanescent wave coupling occurs at the center to fulfill the total internal reflection at the end face of the probe [Illustration at Fig 1 (a)]. The LSPR principle followed by tapered probe is in similitude to that of D-type probe, except the coupling. This probe allows complete TIR without any attenuation as the complete central clad has been removed. Consequently, no dip has been observed. The response with respective change in concentration is depicted in Figure 3 for AuNPs and AgNPs, respectively for better clarity. The sensing parameters are enlisted in Table 1. It is quite evident that in case of tapered probe, the blank response in absence of analyte is quite low as compared to D-shaped fiber. D-shaped fiber exhibits augmented intensity relative to tapered one. However, in terms of intensity profile, tapered probe outsmarts the D-shaped fiber which is evidenced by the appearance of a dip in the latter. The abstinence of dip in tapered one makes the intensity profile better. Meanwhile, in terms of linearity, the D-shaped is superior with respect to the tapered one. In case of tapered one, sensitivity is less with AuNP. On the contrary, the same yields better sensitivity for AgNP relative to D-shaped one. However, D-shaped offers higher limit of detection which is inclusive of both NPs. On the other hand, tapered renders least limit of detection with AuNP as compared to AgNP. The values are considerably smaller than those obtained for D-shaped one. Similarly, both tapered and D-shaped provide substantial working domain in case of AgNPs where as working domain decline in case of AuNPs for both the geometries. Summarily, D-shaped yields better response than its counterpart. All these comparative analysis is elaborated in Table 1.

Table 1: Performance of the proposed set-up

| Noble metal NPs | $R^2$ | | Sensitivity (mV/ppm) | | LOD (ppm) | | Working domain (ppm) | |
|---|---|---|---|---|---|---|---|---|
| | D-type | Tapered | D-type | Tapered | D-type | Tapered | D-type | Tapered |
| AuNPs | 0.999 | 0.993 | 0.0964 | 0.0038 | $174.24 \times 10^{-2}$ | 6.093 | 0-235.37 | 0-437.12 |
| AgNPs | 0.993 | 0.988 | 0.0304 | 0.0039 | $194.75 \times 10^{-2}$ | 13.823 | 0-470.75 | 0-369.88 |

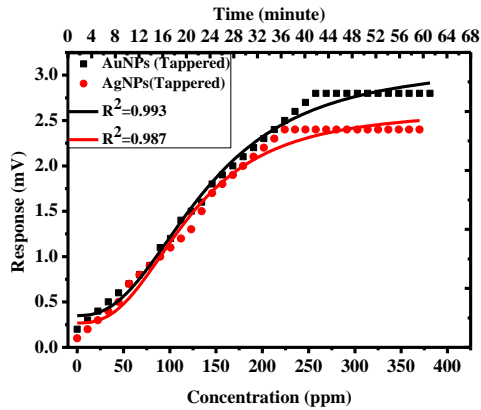

Fig. 3: Response of the sensing set-up vs. concentration of methanol in case of tapered AuNPs and AgNPs coated probe

By fitting the response curve as before, the empirical formula for tapered probe can be expressed as follows for AuNPs and AgNPs, respectively.

AuNPs:
$$Y_m = 3.134 - \frac{2.784}{1 + \left(\frac{X_m}{139.02}\right)^{2.4}} \quad (2a)$$

AgNPs:
$$Y_m = 2.62 - \frac{2.351}{1 + \left(\frac{X_m}{121.46}\right)^{2.6}} \quad (2b)$$

## IV. CONCLUSION

In summary, the sensing performances of two different probe geometries are broadly assessed and corresponding results are analyzed. It is found that the D-type probe is found to show good response towards changes than that of tapered probe. With AgNPs coating, the probe yield regression of ~0.99 and ~0.98. On the other hand, for AuNPs coated probe, the responses are found to be almost similar in each set of configuration. The former provides better responses as compared to AgNPs. The highest sensitivity is found to be ~0.0964 mV/ppm for AuNPs coated D-type optical fiber probe. It is quite apparent that each set of probe is quite responsive towards the plasmonic behavior of the NPs with respect to methanol. However, the tapered fiber also renders an excellent response in terms of intensity. Precisely, each probe can be used for detection of methanol in the range of 0-500ppm. Thus, the proposed sensing set-ups can be used for detection and control of methanol concentration as 200-375ppm range is considered to be harmful for human body inhalation. The simplicity in design due to use of optical detector in lieu of spectrophotometer makes the whole implementation cost-effective and portable. The proposed sensing set-up can also be used to detect other volatile liquids and can also be used to reduce the environment inhalation. With an excellent repeatability rate, these schemes can be extended for real time monitoring with embedded electronics. These schemes have enough potential to be applied for implementation in detection of hazardous vapors in industrial as well as lab safety operations.

REFERENCES

[1] C. Caucheteur, T. Guo and J. Albert, "Review of plasmonic fiber optic biochemical sensor: improving the limit of detection", *Anal. Bioanal. Chem.* vol. 407, pp. 3883-3897, 2015.
[2] Gupta, B., Dodeja, H. and Tomar, A. "Fibre-optic evanescent field absorption sensor based on a U-shaped probe." *Optical and Quantum Electronics*, vol. 28, no. 11, pp. 1629-1639, 1996.
[3] Hsing-Ying Lin, Chen-Han Huang, Gia-Ling Cheng, Nan-Kuang Chen, and Hsiang-Chen Chui, "Tapered optical fiber sensor based on localized surface plasmon resonance", *Optics express*, vol. 20, no. 20, pp. 21693-21701, 2012.
[4] Chiu, M.-H., Wang, S.-F., and Chang, R.-S, "D-type fiber biosensor based on surface-plasmon resonance technology and heterodyne interferometry", *Optics letters*, vol. 30, no. 3, pp. 233-235, 2005.
[5] Mackenzie, H. S. and Payne, F. P., "Evanescent field amplification in a tapered single-mode optical fibre", *Electronics letters*, vol. 26, no. 2, pp. 130-132, 1990.
[6] D. Paul, S. Dutta and R. Biswas, "LSPR enhanced gasoline sensor", *Journal of Physics D: Applied Physics*, vol. 49, pp. 305104, 2016.
[7] D. Paul, S. Dutta, D. Saha and R. Biswas," LSPR based Ultra-sensitive low-cost U-bent optical fiber for volatile liquid sensing", *Sensor and Actuators B (Chemical)*, vol. 250, pp. 198–207, 2017.
[8] D. Paul and R. Biswas, "Highly sensitive LSPR based photonic crystal fiber sensor with embodiment of nanospheres in different

material domain", *Optics and Laser Technology*, vol. 101, pp. 379–387, 2018.
[9] D. Paul and R. Biswas, "Facile fabrication of sensing set-up for size detection of nanoparticles", *IEEE Transactions on Nanotechnology*, 2018. DOI 10.1109/TNANO.2018.2806306
[10] B. S. Baruah, R. Biswas, "Selective detection of arsenic (III) based on colorimetric approach in aqueous medium using functionalised gold nanoparticles unit", *Material Research Express*, 5(1), 2018
[11] B. S. Baruah and R. Biswas, "Localized surface plasmon resonance based U-shaped optical fiber probe for the detection of $Pb^{2+}$ in aqueous medium", *Sensors and Actuators B Chemical*, 2018. doi: 10.1016/j.snb.2018.08.086
[12] B. S. Baruah, R. Biswas, "An optical fiber based surface plasmon resonance technique for sensing of lead ions: A toxic water pollutant", *Optical Fiber Technology*, 46, 2018, 152-156.
[13] B. S. Baruah, R. Biswas and P. Deb, "A green colorimetric approach towards detection of arsenic (III): a pervasive environmental pollutant", *Optics and Laser Technology*, 111, 825-829.
[14] B. S. Baruah, R. Biswas, "Mangifera indica leaf extract mediated gold nanoparticles: a novel platform for sensing of As(III), IEEE Sensors Letter", *IEEE Sensors Letters*, doi: 10.1109/LSENS.2019.2894419
[15] B. S. Baruah, R. Biswas, "Functionalized silver nanoparticles as an effective medium towards trace determination of arsenic (III) in aqueous solution", *Results in Physics*, https://doi.org/10.1016/j.rinp.2019.02.044
[16] Tian, Y., Wang, W., Wu, N., Zou, X., and Wang, X., "Tapered optical fiber sensor for label-free detection of biomolecules", *Sensors*, vol. 11, no. 4, pp. 3780-3790, 2011.
[17] Villatoro, J., Monzon-Hernández, D., and Mejia, E. "Fabrication and modeling of uniform-waist single-mode tapered optical fiber sensors", *Applied Optics*, vol. 42, no. 13, pp. 2278-2283, 2003.
[18] Kieu, K. Q. and Mansuripur, M., "Biconical fiber taper sensors", *IEEE Photonics Technology Letters*, vol. 18, pp. (21/24): 2239, 2006.
[19] Mayer, K. M. and Hafner, J. H., "Localized surface plasmon resonance sensors", *Chemical reviews*, vol. 111, no. 6, pp. 3828-3857, 2011.
[20] Kenneth D. Long, Hojeong Yu, and Brian T. Cunningham. "Smartphone instrument for portable enzyme-linked immunosorbent assays", *Biomedical Opt. Express*, vol. 5, pp. 3793, 2014.
[21] S. Agnihotri, S. Mukherjee and S. Mukherjee, "Size-controlled silver nanoparticles synthesized over the range 5–100 nm using the same protocol and their antibacterial efficacy", *RSC Adv.*, vol. 4, pp. 3974, 2014
[22] Richard N. Cassar, Duncan Graham, Iain Larmour, Alastair W. Wark, Karen Faulds, "Synthesis of size tunable monodispersed silver nanoparticles and the effect of size on SERS enhancement", *Vibrational Spectroscopy*, vol. 71, pp. 41-46, 2014.
[23] Lakshminarayana Polavarapu and Qing-Hua Xu. "A simple method for large scale synthesis of highly monodisperse gold nanoparticles at room temperature and their electron relaxation properties", *Nanotechnology*, vol. 20, pp. 1, 2009.
[24] Sruthi P. Usha, Satyendra K. Mishra and Banshi D. Gupta, "Fabrication and Characterization of a SPR Based Fiber Optic Sensor for the Detection of Chlorine Gas Using Silver and Zinc Oxide", *Materials,* vol. 8, pp. 2204, 2015.
[25] Qiang Zhang, Chenyang Xue, Yanling Yuan, Junyang Lee, Dong Sun and Jijun Xiong," Fiber Surface Modification Technology for Fiber-Optic Localized Surface Plasmon Resonance Biosensors", *Sensors*, vol. 12, pp. 2729-2741, 2012.